\documentclass[aip,graphicx]{revtex4-1}

\usepackage{graphicx}
\usepackage{url}
\usepackage{color}

\newcommand*{\citen}[1]{%
  \begingroup
    \romannumeral-`\x 
    \setcitestyle{numbers}%
    \cite{#1}%
  \endgroup   
}

\draft 

\begin{document}


\title{Measurements of elastoresistance under pressure by combining \textit{in-situ} tunable quasi-uniaxial stress with hydrostatic pressure} 



\author{Elena Gati $^\star$}
\email[]{egati@ameslab.gov}

\altaffiliation{$^\star$ These authors contributed equally to the present work.}
\address{Ames Laboratory, US Department of Energy, Iowa State University, Ames, Iowa 50011, USA}
\address{Department of Physics and Astronomy, Iowa State University, Ames, Iowa 50011, USA}

\author{Li Xiang $^\star$}

\address{Ames Laboratory, US Department of Energy, Iowa State University, Ames, Iowa 50011, USA}
\address{Department of Physics and Astronomy, Iowa State University, Ames, Iowa 50011, USA}

\author{Sergey L. Bud'ko}

\address{Ames Laboratory, US Department of Energy, Iowa State University, Ames, Iowa 50011, USA}
\address{Department of Physics and Astronomy, Iowa State University, Ames, Iowa 50011, USA}

\author{Paul C. Canfield}

\address{Ames Laboratory, US Department of Energy, Iowa State University, Ames, Iowa 50011, USA}
\address{Department of Physics and Astronomy, Iowa State University, Ames, Iowa 50011, USA}

\date{\today}

\begin{abstract}
Uniaxial stress, as well as hydrostatic pressure are often used to tune material properties in condensed matter physics. Here, we present a setup which allows for the study of the combined effects of quasi-uniaxial stress and hydrostatic pressure. Following earlier designs for measurements under finite stress at ambient pressure (e.g., Chu \textit{et al.}, Science \textbf{337}, 710 (2012)), the present setup utilizes a piezoelectric actuator to change stress \textit{in situ} inside the piston-cylinder pressure cell. We show that the actuator can be operated over the full temperature (from 30\,K up to 260\,K) and pressure range (up to $\approx\,$2\,GPa), resulting in a clear and measurable quasi-uniaxial strain. To demonstrate functionality, measurements of the elastoresistance (i.e., the change of resistance of a sample as a response to quasi-uniaxial strain) under finite hydrostatic pressure on the iron-based compound BaFe$_2$As$_2$ are presented as a proof-of-principle example, and discussed in the framework of electronic nematicity. Overall, this work introduces the combination of \textit{in situ} tunable quasi-uniaxial stress and large (up to $\approx$\,2\,GPa) hydrostatic pressure as a powerful combination in the study of novel electronic phases. In addition, it also points towards further technical advancements which can be made in the future.
\end{abstract}


\maketitle 

\section{Introduction}

Optimizing a material's properties by external tuning parameters is of interest to basis physics as well as materials science. From a more fundamental view, systematic control of a material's properties allows for the discovery of exotic phases with novel properties, and forms the experimental basis for developing a theoretical description of these effects and states. In condensed matter physics, the most prominent example of a novel electronic state is superconductivity in various material classes \cite{Keimer15,Paglione10,Mazin10,Canfield10,Steglich16,Kanoda08}, but other examples include non-Fermi liquid behavior \cite{Stewart01,Lee18}, metal-insulator transitions \cite{Lee06,Mott90}, multiferroicity \cite{Hu19} or more recently spin-liquid phases \cite{Balents10,Knolle19} and topological phases \cite{Senthil15}. Often, these states are stabilized by fine-tuning of a material via well-established tuning parameters, such as chemical substitution, magnetic field or pressure, or by a combination of any number of these tuning parameters. 

Combinations of tuning parameters can be particularly powerful, as each tuning parameter is distinct in their effect on the material. For example, non-isovalent substitution affects the bandfilling and thereby modifies the density of states at the Fermi level. In contrast, pressure changes, in the first instance, lattice parameters of the system, which in turn affect the electronic band structure due to electron-lattice coupling. For pressure, the lattice can be perturbed in two different, powerful ways: hydrostatic and uniaxial pressure. Whereas hydrostatic pressure preserves the symmetry of the crystal lattice, uniaxial pressure is directional and therefore can act as a symmetry-breaking field. 

Experimentally, hydrostatic pressure is typically applied by placing a sample into a pressure cell \cite{Eremets96,Fujiwara80,Budko84,Bridgman52,Tozer85} and surrounding it by a pressure-transmitting medium (either gas or liquid). When a force is applied to the medium (e.g., via application of force to a piston), the pressure medium ensures an equal distribution of pressure to all sample surfaces. For uniaxial pressure, there is a range of tools available to apply adjustable pressure. For example, a sample can be fixed between two anvils \cite{Pfleiderer97,Osterman85}, resulting in compressive stress, or tensile stress can be realized experimentally by pulling on an appropiately shaped sample \cite{Cook77,Brandt80}. Recently, other designs have been reported, which utilize voltage-driven piezoelectric actuators \cite{Shayegan03,Chu12,Kuo13,Kuo16,Hicks14,Barber19} to apply stress, and thereby control strain, \textit{in situ} in samples. For this purpose, the samples are either directly attached to the surface of a piezoelectric actuator \cite{Shayegan03,Chu12,Kuo13,Kuo16}, or fixed between two plates, one of which is moved by a piezoelectric actuator \cite{Hicks14,Barber19}.

Here, we present a miniaturization of the piezo-based strain device, as initially presented in Ref. \citen{Chu12}, such that it fits into a conventional piston-pressure cell (with highest pressure of 2\, to 3\,GPa). This new design therefore allows for the study of the combined effects of \textit{in situ}-tunable quasi-uniaxial stress and hydrostatic pressure, i.e., the combination of symmetry-breaking and non-symmetry-breaking tuning parameters. We clearly demonstrate that even up to $p\,\approx\,$2\,GPa, we can induce a, in first approximation linear in applied voltage, strain. Target materials, for which this combination is particularly interesting, include symmetry-broken electronic states of matter, which are coupled to the crystalline lattice. One prominent example examined here is electronic nematic order \cite{Fernandes19}, which is found in various members of iron-based \cite{Fernandes14} or cuprate superconductors \cite{Kivelson98}. In particular for iron-based superconductors, but also other correlated materials \cite{Rosenberg19}, measurements of the elastoresistance were recently established as a tool to investigate nematic fluctuations \citep{Chu12,Kuo16}. Elastoresistance refers to the change of the electrical resistivity as a function of small changes of strain. Here, we will present measurements of the elastoresistance of the iron-pnictide compound BaFe$_2$As$_2$ under pressure, using the combination of uniaxial and hydrostatic pressure, as a proof-of-principle example. These data allow us to to highlight the potential of this tuning combination for future research on novel electronic states, but also to highlight some potential issues with the present design, and outline how these might be improved in the future.

\section{Experimental Setup}
\label{sec:setup}

\begin{center}
	\begin{figure}
	\includegraphics[width=0.7\textwidth]{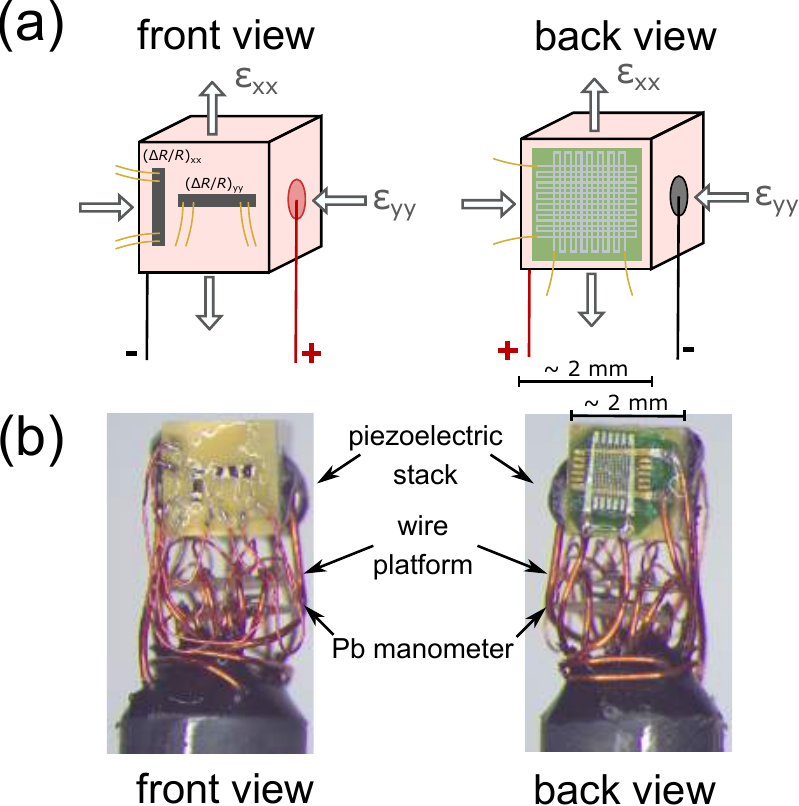}
	\caption{(a) Schematic view of the piezoelectric actuator, which is used to apply uniaxial strain to the samples (front view is shown on the left, back view on the right). Dark grey arrows indicate the longitudinal and transverse strains, $\epsilon_{xx}$ and $\epsilon_{yy}$, respectively, which are induced by applying a voltage to the piezoelectric actuator (depicted by ''+'' and ''-''). Black bars in the front view represent two samples, which are fixed to the piezoelectric actuator and prepared for resistance measurements, using a standard four-point technique. The green rectangle in the back view represents the crossed strain gauges, which allow for an \textit{in situ} determination of $\epsilon_{xx}$ and $\epsilon_{yy}$ during a voltage sweep; (b) Photograph of the piezoelectric actuator mounted on the pressure-cell feedthrough. In addition, two platforms are placed between the piezoelectric actuator and the feedthrough; the upper platform is used to connect the sample wires to the feedthrough and allows a fast mounting and removing of the sample from the feedthrough. On the second platform, a lead (Pb) manometer for \textit{in situ} pressure determination is mounted.}
	\label{fig:schematics} 
	\end{figure}
	\end{center}

To vary stress/strain \textit{in situ}, a positive bias voltage is applied to a commercial piezoelectric actuator (see Fig.\,\ref{fig:schematics} (a)), as a result of which the actuator elongates along the stacking direction of the piezoelectric layers (denoted as strain $\epsilon_{xx}$) and shrinks perpendicular to it ($\epsilon_{yy}$). For the present experimental design, the maximum size of the piezoelectric actuator for applications under hydrostatic pressure is determined by the inner diameter of the sample space inside the pressure cell (typically $\,\approx\,3.2\,$mm). We therefore chose a piezoelectric actuator from Thorlabs (Item \#PA4CE) with dimensions of $2\,\times\,2\,\times\,2\,$mm$^3$. For the given drive voltage range of 0\,V to 150\,V, the maximum displacement along the stacking direction is $2\,\mu$m (corresponding to a strain of 0.1\,\%). Thin samples are attached to the surface of the actuator to study their response to the quasi-uniaxial strain, generated by the piezoelectric actuator. Strictly speaking, any measurement, in which a sample is attached to a piezoelectric actuator, is performed under biaxial stress \cite{Kuo13}, as the actuator evidently changes its dimensions in longitudinal and transverse direction, and as such, a sample will also be strained in both directions. However, the strain in both directions is highly anisotropic, as it is opposite in sign, and is therefore still highly directional (similar to uniaxial strain). We thus refer to these conditions as quasi-uniaxial stress/strain throughout the manuscript.

So as to provide electrical connections to the piezoelectric actuator, to which samples are attached and which is subsequently inserted into the pressure cell, a home-built pressure cell feedthrough was used. For this particular case, the feedthrough was equipped with a large number of wires (up to 20 Cu wires; see Fig.\,1(b) for a photograph of the feedthrough), which are used to supply the driving voltage for the piezo, for measurements of the samples' resistance changes as a response to the external strain as well as for the \textit{in situ} determination of the strain, created by the actuator, and of the pressure inside the pressure cell. In the following, the individual components each will be explained  separately.

The two silver-coated electrodes of the piezoelectric actuator are soldered directly to two of the feedthrough wires, which were chosen to be thick enough ($>\,100\,\mu$m) so as to also provide sufficient mechanical stiffness to support the actuator itself. Given the high voltages, which are needed to drive the actuator, it has to be ensured that these wires are electrically well-insulated from their environment so as to avoid a voltage breakdown and/or a leakage of voltage. Tests of our setup up to the maximum applied voltage of 150\,V did not show any signatures of a breakdown or leakage; the pressure medium used in the present case (a 4:6 mixture of light oil and $n$-pentane), as well as the Stycast epoxy (2850FT), which is used to secure the wires and also, thereby, seal the feedthrough, provide a sufficient insulating environment. It is worth mentioning that, during tests, the usage of a different pressure medium with higher conductivity (60:40 glycerol-water mixture) resulted in a leakage of the voltage applied to the actuator to the conducting samples.

To attach thin samples to the side of the piezoelectric actuator, a two-component epoxy (Devcon 14250, General Purpose Adhesive Epoxy) is used. Typically, a single sample is sufficient to study the combined response of a sample to uniaxial and hydrostatic lattice deformations. However, here, for reasons outlined below in Sec. \ref{sec:results-Ba122}, two samples are used in the present study. They are placed orthogonally on the actuator (shown schematically in Fig.\,\ref{fig:schematics}\,(a)). Each sample is contacted in a standard, linear four-point configuration for resistance measurements, by spot-welding Pt wires to a cleaved surface of a thin crystal and secured by a drop of silver-paint (Dupont 4929N). These samples are typically glued to the side of the actuator, after the actuator itself is fixed on the feedthrough. To guarantee sufficient mechanical stability of the electrical connections between sample wires and feedthrough, and to also allow for a fast mounting and removing of the sample from the feedthrough, the sample wires (via thin Cu wires) and the feedthrough wires were connected on a platform, which was mounted underneath the actuator (see ''wire platform'' in Fig.\,\ref{fig:schematics}\,(b)).

For an accurate \textit{in situ} determination of the strain generated by the piezoelectric actuator, a strain gauge is attached to the opposite side of the actuator, again using the two-component epoxy from Devcon (Devcon 14250, General Purpose Adhesive Epoxy). Strain gauges are sensors, which utilize changes of their resistance as a result of changes of their geometry for the determination of the strain along a specific direction via a known gauge factor $K\,\approx\,2$. To simultaneously measure $\epsilon_{xx}$ and $\epsilon_{yy}$, a two-element strain gauge (Type FCA-1-23, Tokyo Measuring Instruments Lab.) is used, where two orthogonal strain gauges are stacked in one coating (see Fig.\,\ref{fig:schematics}, right). Each of these strain gauges (with typical resistances of $\approx\,120\,\Omega$ and gauge dimension of 1\,mm) are connected to two feedthrough wires inside the pressure cell, but measured in a four-point configuration outside of the pressure cell.

In addition, the pressure inside the pressure cell is determined via the shift of the superconducting transition temperature, $T_c$, of elemental lead (Pb) \cite{Bireckoven88}, which is mounted on a separate platform below the wire platform (see Fig.\,\ref{fig:schematics} (b)) in a four-point configuration for resistance measurements. The so-inferred pressure values correspond to the pressure at low temperatures, which typically increases as the temperature is increased. In the present piston-cylinder cell, this increase could be as high as $\,\approx\,0.3\,$GPa upon reaching room temperature. Throughout the entire manuscript, pressure values correspond to the measured, low-temperature pressure values. Further, the label ''$p\,=\,0\,$GPa'' refers to the situation, in which the sample is placed inside the pressure cell and which was closed ''hand-tight'' at room temperature. This procedure typically gives rise to a very small pressure (close to ambient) at low temperatures, as confirmed by the Pb manometer. Some measurements (for comparison) were also performed outside of the pressure cell in true ambient pressure conditions. These are then labeled with ''ambient'' or ''outside of the pressure cell''.

	The sample end of the feedthrough is placed in a Teflon-cup filled with the pressure medium. For all experiments shown, a 4:6 volumetric mixture of light oil and $n$-pentane was used. As mentioned above, it does not only guarantee sufficient electrical insulation between all wires, but also provides good hydrostatic pressure conditions \cite{Kim11,Torikachvili15} up to highest pressures of 2\,GPa to 3\,GPa, since solidification of this medium takes place above 3\,GPa at room temperature. Anti-extrusion rings, machined out of phosphor bronze, are used to prevent Teflon from flowing through interstices when pressurized. The pressure cell, used for all experiments, is a double-wall cell with the outer cylinder made out of Grade 5 titanium alloy (Ti 6Al-4V) and the inner cylinder out of Ni-Cr-Al alloy (see Ref. \citen{Budko84} for a very similar design). 
	  	
	 The cryogenic environment was provided by a closed-cycle cryostat (Janis SHI-950 with a base temperature of $\approx\,3.5$\,K). The probe was equipped with phosphor-bronze wires (QT-36, LakeShore, Inc.) to ensure low heat flow through the wires, which becomes particularly important in the present case, where a large number of wires are needed. Temperature was controlled using a LakeShore 336 controller and monitored by using a calibrated temperature sensor (Cernox-1030) which was placed directly on the outer surface of the pressure cell. In the present study, voltage sweeps were performed at constant temperature. To ensure good thermal equilibrium of the samples inside the pressure cell, each temperature was stabilized and held for 15 minutes prior to the voltage sweep. To check that that the samples inside of the pressure cell are indeed sufficiently thermalized, the resistance of the sample served as a good reference, as it is very sensitive to changes with temperature. After the 15 minutes hold time, no change of the sample resistance was resolved for at least five minutes. Data were taken for $T\,<\,260\,$K, since earlier works \cite{Chu12,Kuo16} demonstrated that, for higher temperatures, and this particular glue, strain is only poorly transmitted to the sample ($\epsilon_{sample}<\,80\,\%\,\epsilon_{piezo}$). In addition, data were not taken below 30\,K, since these temperatures are well below the temperature region of interest for BaFe$_2$As$_2$, studied here. In general, the operation of the actuator, however, is not expected to be limited to 30\,K.
	 
	 A voltage source (E3640A, Agilent) together with an amplifier (Analog Amplifier SVR 150/1, Piezomechnanik GmbH) was used to apply voltage to the piezoelectric actuator. For the measurements shown, voltage was swept from 0\,V to 150\,V, and back to 0\,V at a rate of $\,\pm\,$3\,V/s. The sample resistances and strain gauge resistances were recorded simultaneously during each voltage sweep. In total, three sweeps were performed at each temperature to ensure reproducibility. The sample resistances, as well as the resistance of the Pb manometer were measured with LakeShore AC resistance bridges (Models 370 and 372). The strain gauge resistances were measured using Keithley 2001 and 2010 multimeters. Data were recorded by a customized LabView program.

\section{Results}
\label{sec:results}

	\subsection{Generated strain by the piezoelectric actuator under pressure}
	\label{sec:straintransmission}
	
	First, we demonstrate that a piezoelectric actuator, which is placed inside a pressure cell, can be strained by an applied voltage over wide ranges of temperature (low $T\,\approx$\,30\,K up to 260\,K) and pressures (0\,GPa$\,<\,p\,\lesssim\,2$\,GPa). This result cannot be expected \textit{a priori}, as (i) the piezoelectric actuator might break, if exposed to large pressures or (ii) it might not be able to act against the significant external forces, it is exposed to. In fact, no breakage or significant damage was observed in visual inspection of the actuator after a pressure cycle. In what follows, we present a detailed characterization of the piezoelectric actuator under pressure, with focus on the longitudinal and transverse strain for different temperatures and pressures, measured by the orthogonal strain gauges. 
	
\begin{center}
	\begin{figure}
	\includegraphics[width=0.95\textwidth]{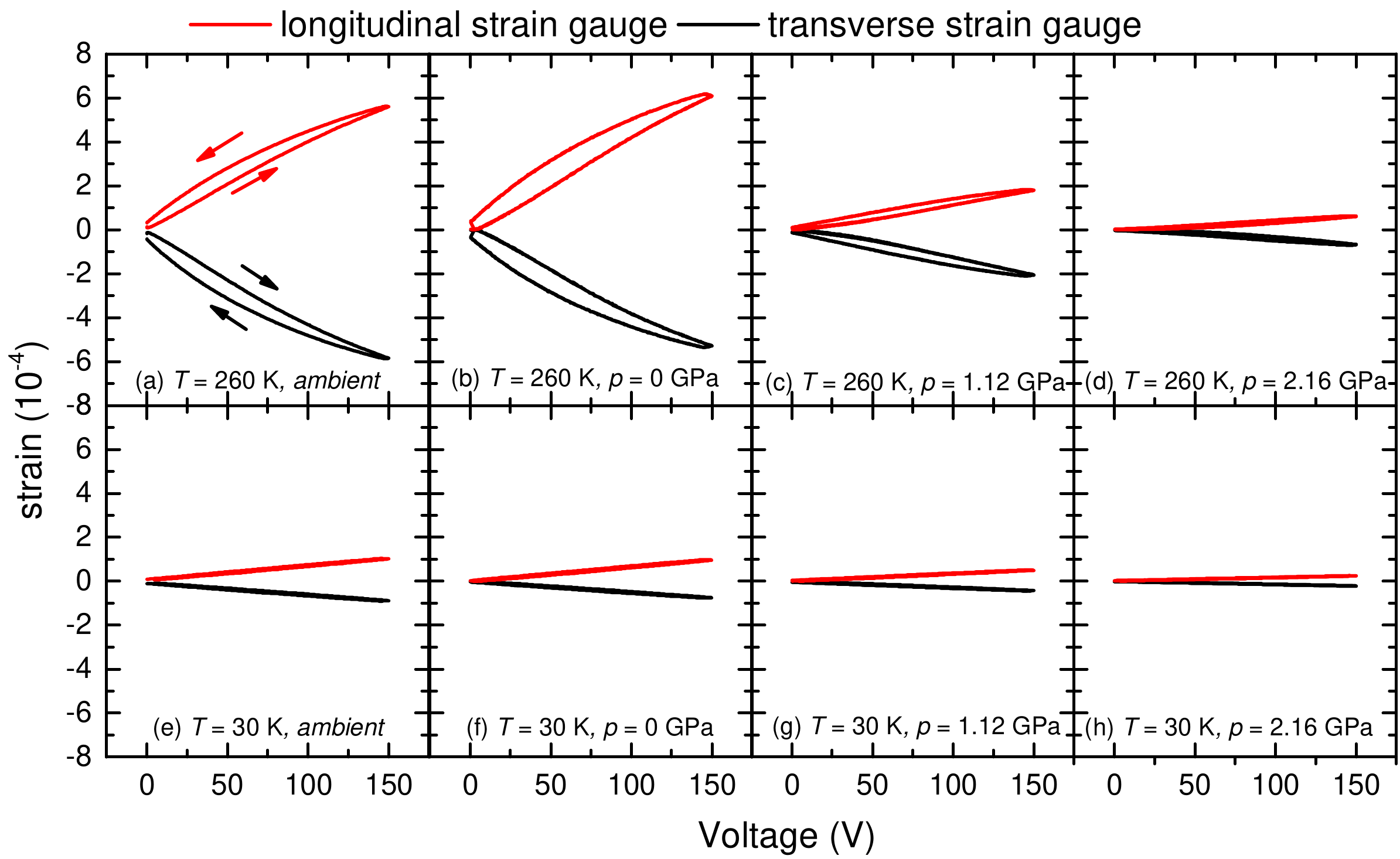}
	\caption{Longitudinal and transverse strain, induced by applying a voltage to a piezoelectric stack inside the pressure cell and measured by an orthogonal two-element strain gauge \textit{in situ}, at ambient pressure outside of the pressure cell (a,e), at $p\,=\,0\,$GPa inside the pressure cell (b,f), 1.12\,GPa (c,g) and 2.16\,GPa (d,h) at high temperature $T\,=\,260\,$K (top) and low $T\,=\,30\,$K (bottom). Arrows in (a) indicate data taken upon increasing and decreasing voltage.}
	\label{fig:transmittedstrain} 
	\end{figure}
	\end{center}

	In general, strain $\epsilon$ is defined as relative change of length $l$ along a particular direction $i$, $\epsilon\,=\,\frac{l_i-l_{i,0}}{l_{i,0}}$, with $l_{i,0}$ being the unstrained length. Throughout the entire manuscript, we will use the notation of positive strain ($\epsilon\,>$\,0) for elongation of the actuator along a particular direction, while negative strain ($\epsilon\,<$\,0) denotes compression along a particular direction. In Fig.\,\ref{fig:transmittedstrain}, we present example data sets of longitudinal and transverse strain as a function of applied voltage for ambient pressure outside the cell (a,e), low pressure inside the cell ($p\,=\,0$\,GPa, b,f), medium pressure (1.12\,GPa, c,g) and high pressure (2.16\,GPa, d,h), each for a high temperature ($T\,=\,260\,$K, top) and a low temperature (30\,K, bottom).  The key result here is that for any of these combinations of pressure and temperature, the application of voltage results in clear and measurable longitudinal and transverse length changes of the piezoelectric actuator. The observed hysteresis between increasing and decreasing voltage is characteristic for any piezoelectric material (see, e.g., arrows in Fig.\,\ref{fig:transmittedstrain}). We note that at $T\,=\,30\,$K, the pressure medium is solid for all pressures shown here (see Fig.\,\ref{fig:transmittedstrain} (f-h)). However, compared to the data set at 30\,K outside of the pressure cell (see Fig.\,\ref{fig:transmittedstrain} (e)), the strain-voltage characteristic is nearly unchanged for lowest pressure ($p\,=\,0$\,GPa) inside the cell. This therefore demonstrates that the solidification of the medium does not significantly compromise the operation of the actuator.
	
	In more detail, the strain generated by a voltage of 150\,V, $\epsilon_{150\,V}$, is largest for highest temperature and lowest pressure, for the longitudinal as well as the transverse direction; $\epsilon_{150\,V}$ decreases either with decreasing temperature, or with increasing pressure. The decrease of the displacement with temperature is well-known for piezoelectric actuators; it might be possible to compensate for this decrease by operating the actuator to higher voltages (also against its poling direction) \cite{Hicks14}, as the coercive field strength typically increases, upon cooling below the Curie temperature of the piezoelectric material. However, no voltages higher than 150\,V, following the actuator's room-temperature specifications, were applied in the present case. Nonetheless, this might be an option in the future to increase the amount of strain at low temperatures. The decrease of $\epsilon_{150\,V}$ with pressure agrees well with the naive expectation that the actuator has to counteract the increasing force, exerted by the pressure medium (provided that the work, which the actuator can perform, remains constant). A simple extrapolation of the longitudinal $\epsilon_{150\,V}$ to zero yields a critical pressure of $\approx\,(3\,\pm\,0.5)$\,GPa for the operation of the present actuator.

		A comparison between displacements in longitudinal and transverse directions shows that, on a quantitative level, $\epsilon_{xx}$ and $\epsilon_{yy}$ are opposite in sign, but of similar magnitude, for all pressures and temperatures. A quantitative evaluation of $\epsilon_{xx}$(150\,V) and $-\,\epsilon_{yy}$(150\,V) for $30\,$K$\,\le\,T\,\le\,$260\,K and $0\,$GPa$\,\le\,p\,\le\,$2.16\,GPa is summarized in Fig.\,3, for one of the actuators. Overall, only small variations in the size of $\epsilon_{xx}$ and $\epsilon_{yy}$ were observed ($\frac{\Delta \epsilon_{xx}}{\epsilon_{xx}}\,\simeq\,\frac{\Delta \epsilon_{yy}}{\epsilon_{y}}\,\lesssim\,5$\,\%) for different actuators of the same type, which thus behave very similarly on a semi-quantitative level. 
	
\begin{center}
	\begin{figure}
	\includegraphics[width=0.9\textwidth]{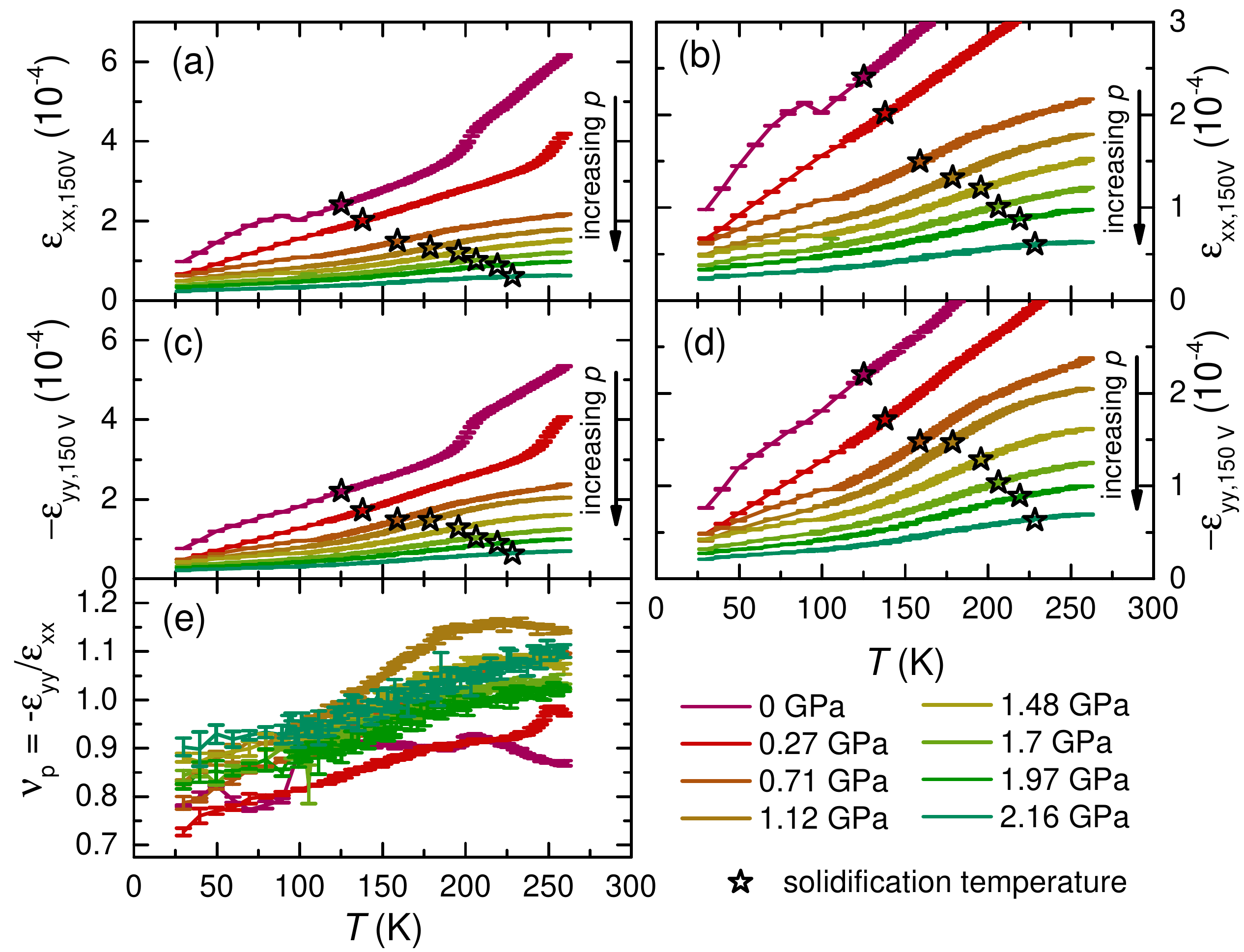}
	\caption{Longitudinal strain, $\epsilon_{xx,150 V}$ (a), transverse strain, $\epsilon_{yy,150 V}$ (c), (each induced by a voltage of 150\,V), and Poisson ratio, $\nu_p\,=\,-\frac{\epsilon_{yy}}{\epsilon_{xx}}$, (e) of the piezoelectric actuator as a function of temperature, $T$, for different pressures between 0\,GPa and 2.16\,GPa. The right plots in the top and middle panel ((b) and (d)) show data on the left on enlarged scales. Black stars mark the pressure-dependent solidification temperature of the pressure medium used \cite{Torikachvili15}.}
	\label{fig:straintemperaturepressure} 
	\end{figure}
	\end{center}

	As shown in the top and middle panels of Fig.\,\ref{fig:straintemperaturepressure} (denoted as (a)-(d)), $\epsilon_{xx, 150 V}$ and $-\epsilon_{yy, 150 V}$ change mostly monotonically as a function of $T$ and $p$ for $p\,\gtrsim\,0.3\,$GPa. The reason for sudden changes of $\epsilon_{xx, 150 V}$ and $-\epsilon_{yy, 150 V}$, particularly pronounced for $p\,=\,0\,$GPa at $T\,\approx\,200\,$K and 100\,K and for 0.27\,GPa at $\approx\,$250\,K, is unclear at present. Similar effects were observed for other piezoelectric actuators at ambient pressure after moderate usage \cite{Kuo14}. Importantly, to a good approximation, this should not affect the analysis of the sample's response, presented below, as strain is measured \textit{in situ} for all temperatures and pressures.

	For $p\,=\,$0\,GPa, the low-temperature displacement corresponds to $\,\approx\,15\,\%$ of the high-temperature displacement, whereas at highest pressure ($p\,=\,2.16\,$GPa) the low-temperature strain is $\approx\,$65\,\%, of the high-temperature value (at this pressure). Correspondingly, in the most extreme case of lowest temperature and high pressure, the strain corresponds to $\approx\,$4\,\% of the high-temperature, low-pressure strain value. Thus, since this value is not significantly lower than the temperature-induced reduction at ambient pressure and this piezo-based technique has proven to be powerful at ambient pressure, these piezoelectric actuators hold great promise to create sufficient strain to observe a response of the sample. Again, it is interesting to note that no pronounced feature can be associated with the solidification of the pressure medium \cite{Torikachvili15}, which takes place around $T\,\approx\,120$\,K at ambient pressure and increases in temperature up to $\approx\,220\,$K for 2\,GPa for the light mineral-oil/$n$-pentane medium used (see stars in Fig.\,\ref{fig:straintemperaturepressure}).
	
The anisotropy in the displacement of the actuator along xx- and yy-direction is characterized by the so-called Poisson ratio $\nu_p\,=\,-\epsilon_{yy}/\epsilon_{xx}$. The temperature- and pressure dependence of $\nu_p$ is shown in the bottom panel of Fig.\,\ref{fig:straintemperaturepressure} (e). As mentioned earlier, for the present actuator, $\nu_p$ is close to unity with only weak temperature- and pressure dependence. In the next section, we will outline the relevance of the Poisson ratio to the data analysis of elastoresistance. At this point, it is only important that $\nu_p$ is determined \textit{in situ} for any pressure and temperature (via measurements of $\epsilon_{xx}$ and $\epsilon_{yy}$), and therefore can be readily used as an input parameter in the data analysis.

	\subsection{Test case: Elastoresistance of BaFe$_2$As$_2$ under pressure}
	\label{sec:results-Ba122}
	
	In this section, we utilize the combination of quasi-uniaxial stress and hydrostatic pressure to study nematicity in BaFe$_2$As$_2$ under pressure. For this purpose, we first provide a short introduction into nematic order in iron-based superconductors, and why measurements of elastoresistance (i.e., relative change of resistance as a function of strain) provide a useful tool to probe nematic fluctuations (see also Refs. \citen{Chu12,Kuo16}). Afterwards, we turn to our results, which serve to demonstrate the functionality of this setup.
	
	\subsubsection{Elastoresistance as a probe for nematicity}
	\label{sec:results-Ba122-intro}
	
	The emergence of superconductivity in many high-temperature superconductors, such as the iron-based or cuprate superconductors, is generally associated with fluctuations of their unusual normal state, out of which superconductivity is born. For iron-based superconductors, the normal state is, in many cases antiferromagnetic, as is the case for the prototypical iron-based superconducting system\cite{Canfield10} Ba(Fe$_{1-x}$Co$_x$)$_2$As$_2$. The transition into the magnetic state is often preceded by a structural transition from a tetragonal to an orthorhombic crystal structure. In some other cases, like FeSe \cite{Boehmer17}, only the structural transition, but no subsequent magnetic order is found at ambient pressure. Nowadays, there is a common understanding that the structural transition is not simply driven by lattice degrees of freedom, but driven by a symmetry-broken electronic state \cite{Chu12,Fernandes14}. In analogy to the terminology used in liquid crystals, this state is therefore referred to as ''nematic'' state. Experiments suggest that a nematic quantum critical point might exist close to optimal doping \cite{Kuo16,Suguru16,Hong19,Straquadine19} (i.e., close to the concentration at which maximal superconducting transition temperature is observed), in accordance with recent theories \cite{Lederer17}.
	
	Measurements of the elastoresistance in the tetragonal state, in particular on Ba(Fe$_{1-x}$Co$_x$)$_2$As$_2$, have contributed substantially to this understanding of the nematic state: the coefficient of elastoresistance along the [110]$_T$ direction of the tetragonal lattice (which is rotated by 45$^\circ$ from the orthogonal unit cell direction, and therefore corresponds to the direction of orthorhombic distortion), $m_{66}$, diverges upon approaching the nematic transition from above in the tetragonal state \cite{Kuo16}. This result can only be rationalized, if the structural distortion itself is not the primary order parameter of the nematic transition, but rather occurs as a result of the coupling of the electronic nematic state to the crystal lattice \cite{Kuo13}. 
	
	In the following, we will review some key concepts of measurements of elastoresistance for tetragonal materials (such as the iron-based superconductors). Detailed derivations of the formulas can be found in the works which introduced the methodology, e.g. Refs. \citen{Kuo13,Kuo16}. The quantity, which serves as a proxy for the nematic order parameter, is the in-plane resistivity anisotropy, $N$, defined as $N\,=\,\frac{\rho_{xx}-\rho_{yy}}{\frac{1}{2}(\rho_{xx}+\rho_{yy})}$, with $\rho_{xx}$ and $\rho_{yy}$ the resistivity in $xx$ and $yy$ direction, respectively when strain is applied along the tetragonal [110]$_T$ direction (i.e., in Fig.\,\ref{fig:schematics}\,(a) front view, the two sample bars are cut along the [110]$_T$ direction). To the leading order, $N$ can be expressed as
	
	\begin{equation}
	N \,=\, [(\Delta R/R)_{xx}-(\Delta R/R)_{yy}],
	\end{equation}

	with $R_{xx}$ ($R_{yy}$) being the resistance of the sample with the long axis parallel to $\epsilon_{xx}$ ($\epsilon_{yy}$) (see Fig.\,\ref{fig:schematics}).	The resistance change in longitudinal direction, $(\Delta R/R)_{xx}$, for a tetragonal material, can be written as
	
	\begin{equation}
	(\Delta R/R)_{xx} = m_{11} \epsilon_{xx} + m_{12} \epsilon_{yy} + m_{13} \epsilon_{zz},
	\end{equation}
	
	with $m_{ij}$ being the coefficients of the elastoresistance tensor, and $\epsilon_{ii}$ being strains in xx-, yy- and zz-directions. Here, it becomes important that measurements are, in fact, performed under strongly anisotropic biaxial strain, rather than uniaxial strain. As a consequence, $\epsilon_{yy}$ is determined by the Poisson's ratio of the piezoelectric actuator ($\nu_p$, measured experimentally, see Sec. \ref{sec:straintransmission}). The last term, given by $\epsilon_{zz}$, is determined by the Poisson ratio of the sample in $c$ direction, $\nu_s$. For a strain applied along the [110]$_{T}$ direction of the sample, the change of the sample resistance is
	
	\begin{equation}
	(\Delta R/R)_{xx} = \epsilon_{xx} \{\frac{1}{2}(m_{11}+m_{12}+2m_{66})-\nu_p[\frac{1}{2}(m_{11}+m_{12}-2m_{66})]-\nu_s m_{13}\}.
	\end{equation}
	
	This equation can be simplified under the assumption that $\nu_p\,\approx\,1$ (as indeed the case for the present actuator, in particular above $T\,\approx\,100\,$K), to
	
	\begin{equation}
	(\Delta R/R)_{xx}\,\simeq\,\epsilon_{xx} (2 m_{66} - \nu_s m_{13}),
	\end{equation}
	
	and similarly for the transverse direction, $(\Delta R/R)_{yy}\,\simeq\,\epsilon_{xx} (-2 m_{66} - \nu_s m_{13})$. Thus, subtraction of the longitudinal and transverse response (measured on two different samples) yields a full symmetry decomposition (even if $v_p\,\neq\,1$), and the resulting anisotropy $N/2 = -2 m_{66} \epsilon_{xx}\,$ solely depends on the ''nematic'' susceptibility $m_{66}$.
	
	It should be noted, though, that in practice the subtraction of the response of two individual samples might turn out to be complicated, in particular for higher pressures. The reason for this is that the subtraction of the response of two samples with different strain homogeneity (due to different gluing and different thermal expansion of the piezoelectric actuator in different directions) might yield significant errors. Nonetheless, $(\Delta R/R)_{xx}$ or $(\Delta R/R)_{yy}$ themselves can be used to infer the nematic susceptibility in the limit $2m_{66}\,\gg\,\nu_s m_{13}$ (and $\nu_p\,\approx\,$1), which then results in $(\Delta R/R)_{yy}\,\simeq\,\epsilon_{xx} (-2 m_{66})\,\simeq\,-m_{66}(\epsilon_{xx}-\epsilon_{yy})$. For the iron-based superconductors, discussed here, this limit is indeed the case, as $m_{13}$ is typically small and not strongly temperature-dependent $(m_{13}\,\ll\,2m_{66})$ \cite{Kuo13,Palmstrom17,Straquadine19}, and $\nu_s$ is typically of the order of one \cite{Straquadine19}. Therefore, whereas for lowest pressures, we show data of $m_{66}$ (obtained by subtracting d$(\Delta R/R)_{xx}$/d$(\epsilon_{xx}-\epsilon_{yy})$ and d$(\Delta R/R)_{yy}$/d$(\epsilon_{xx}-\epsilon_{yy})$ signals) for comparison with literature results. For higher pressures we focus on the analysis of d$(\Delta R/R)_{yy}$/d$(\epsilon_{xx}-\epsilon_{yy})$ solely to discuss the temperature dependence of $m_{66}$.

	\subsubsection{Experimental Results}
	
	Now we turn to our experimental results of elastoresistance on the iron-pnictide compound BaFe$_2$As$_2$ under hydrostatic pressure. The single crystals used for the present study were grown from flux growth using self-flux, as reported elsewhere \cite{Ni08}. The crystals were cleaved, and cut into two bar-shaped pieces, with the long axis corresponding to the tetragonal [110] direction. Typically, these crystals had dimensions of $0.9\,\times\,0.2\,\times\,0.1\,$mm$^3$. On the one hand, the small thickness along the $c$ axis is very important, so that the samples are strained as homogeneously as possible. On the other hand, given that thin samples are attached directly to the surface of the piezo, this gives rise to two issues, which are unavoidable in the present design: First, due to the non-negligible thermal expansion of the piezoelectric actuator, the samples are already mildly strained without the application of an external voltage to the actuator. However, due to the opportunity of changing strain \textit{in situ}, we can monitor \textit{small changes} of resistance as a result of \textit{small changes} of strain, as long as the sample is in the linear regime. Thus, we can access the nematic susceptibility $\chi_{nem}\,=\,\frac{\partial N}{\partial \epsilon}$, which is the quantity of interest here. Second, only the top surface of the crystal is clearly exposed to the pressure medium in the present design, bringing up the question of the extent of true hydrostaticity of the applied pressure. We cannot rule out significant non-hydrostatic pressure components in this design, but we will show below that the change of transition temperatures with pressure, determined in this work, is similar to the one of free-standing samples. In the future, certainly, one can extend the present design such, that the sample is not entirely attached to the actuator (by e.g. milling a slit into the actuator just below the sample position or mounting the sample between two bars that are attached to the actuator). This should allow the pressure medium to surround most of the sample and therefore improve hydrostaticity significantly, without loosing the ability to strain the sample. As of now, the present data on BaFe$_2$As$_2$ should be thought of as a demonstration of the principle idea of using piezoelectric actuators inside the pressure cell to measure elastoresistance under pressure.
	
\begin{center}
	\begin{figure}
	\includegraphics[width=\textwidth]{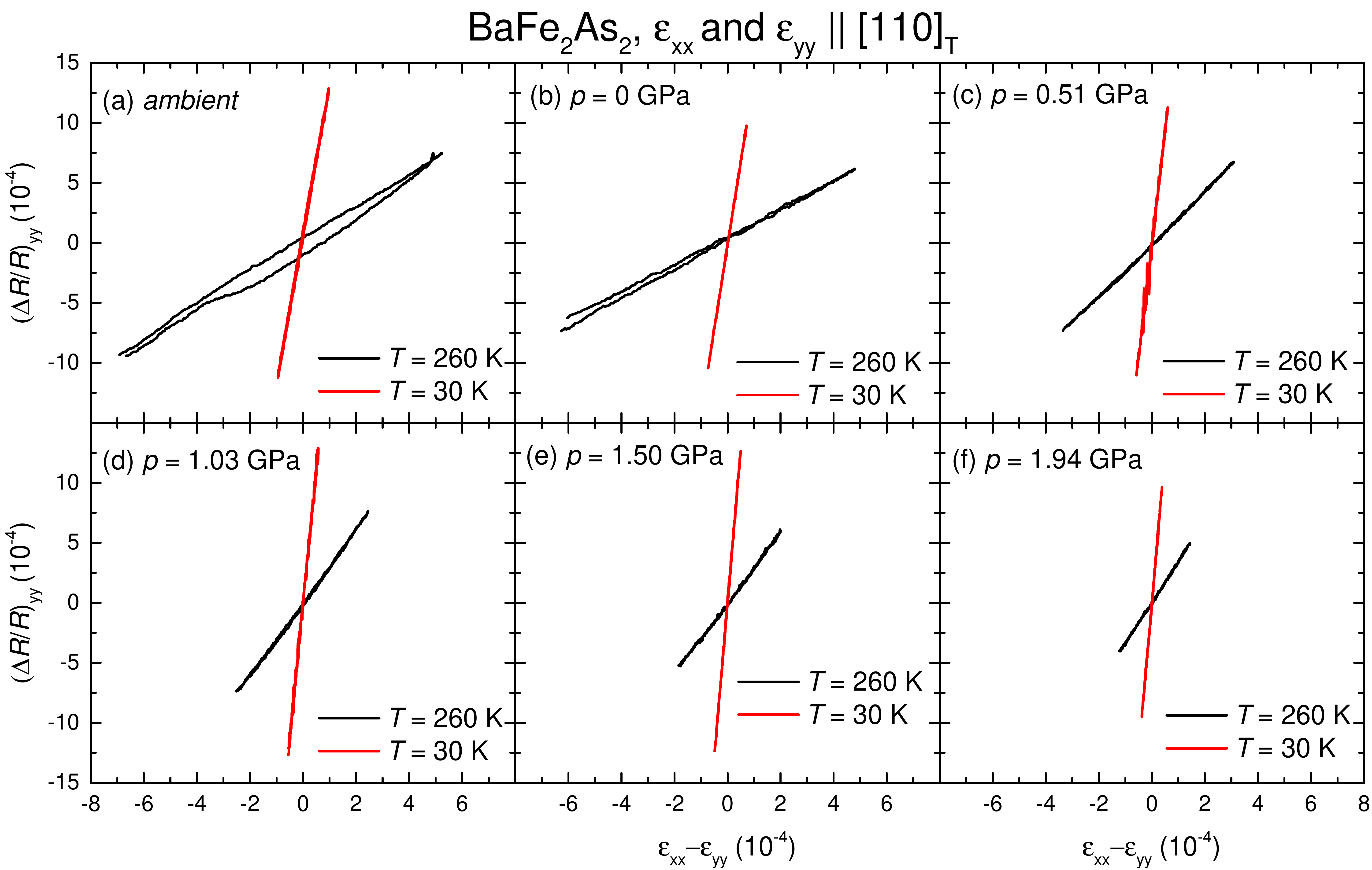}
	\caption{Change of transverse resistance, $(\Delta R/R)_{yy}$, of BaFe$_2$As$_2$ as a function of strain, $\epsilon_{xx}-\epsilon_{yy}$, at $T$\,=\,260\,K and $T$\,=\,30\,K for  ambient pressure outside the pressure cell and five different pressures between 0\,GPa and 1.94\,GPa inside the pressure cell. For these measurements, the strain $\epsilon_{xx}$ and $\epsilon_{yy}$ are applied along the [110] direction of the tetragonal unit cell (see text for more details). Measurements were performed in a cycle, in which voltage was first increased to the maximum value (150 V) and then decreased to 0 V, giving rise to increasing and decreasing strain, respectively.}
	\label{fig:Ba122-Rvseps} 
	\end{figure}
	\end{center}

	In Fig.\,\ref{fig:Ba122-Rvseps}, we show the $(\Delta R/R)_{yy}$ data for of a BaFe$_2$As$_2$ sample, oriented with the [110]$_T$ direction along the strain direction, as a function of $\epsilon_{xx}-\epsilon_{yy}$ for ambient pressure outside the pressure cell (a), as well as for five different pressures ($0\,$GPa$\,\le\,p\,\le\,1.94$\,GPa, (b)-(f)). For each pressure, data at two different temperatures ($T\,=\,30\,$K and $T\,=\,260\,$K) are displayed. In each case, normalization (due to the unknown $R_{yy}(\epsilon = 0)$ as a result of thermal expansion mismatch between actuator and sample) was performed such that the $(\Delta R/R)_{yy}$ vs. $\epsilon_{xx}-\epsilon_{yy}$ is symmetric around zero. Clearly, a change of resistance is detected for all temperature/pressure combinations. Again, we want to stress that this includes ranges of temperatures and pressures, for which the generated strain is reduced due to (i) low temperatures and/or (ii) large force on the actuator resulting from high pressure.  In addition, for each pressure, $(\Delta R/R)_{yy}$ changes in a linear fashion with $\epsilon_{xx}-\epsilon_{yy}$, without any significant hysteresis between increasing and decreasing strain. This indicates that we probe the response of the sample in the linear regime for all pressures and temperatures. Also, it is evident that, for each pressure, the response of BaFe$_2$As$_2$ to strain is larger at $T\,=\,$30\,K, compared to the one at 260\,K.

\begin{center}
	\begin{figure}
	\includegraphics[width=0.5\textwidth]{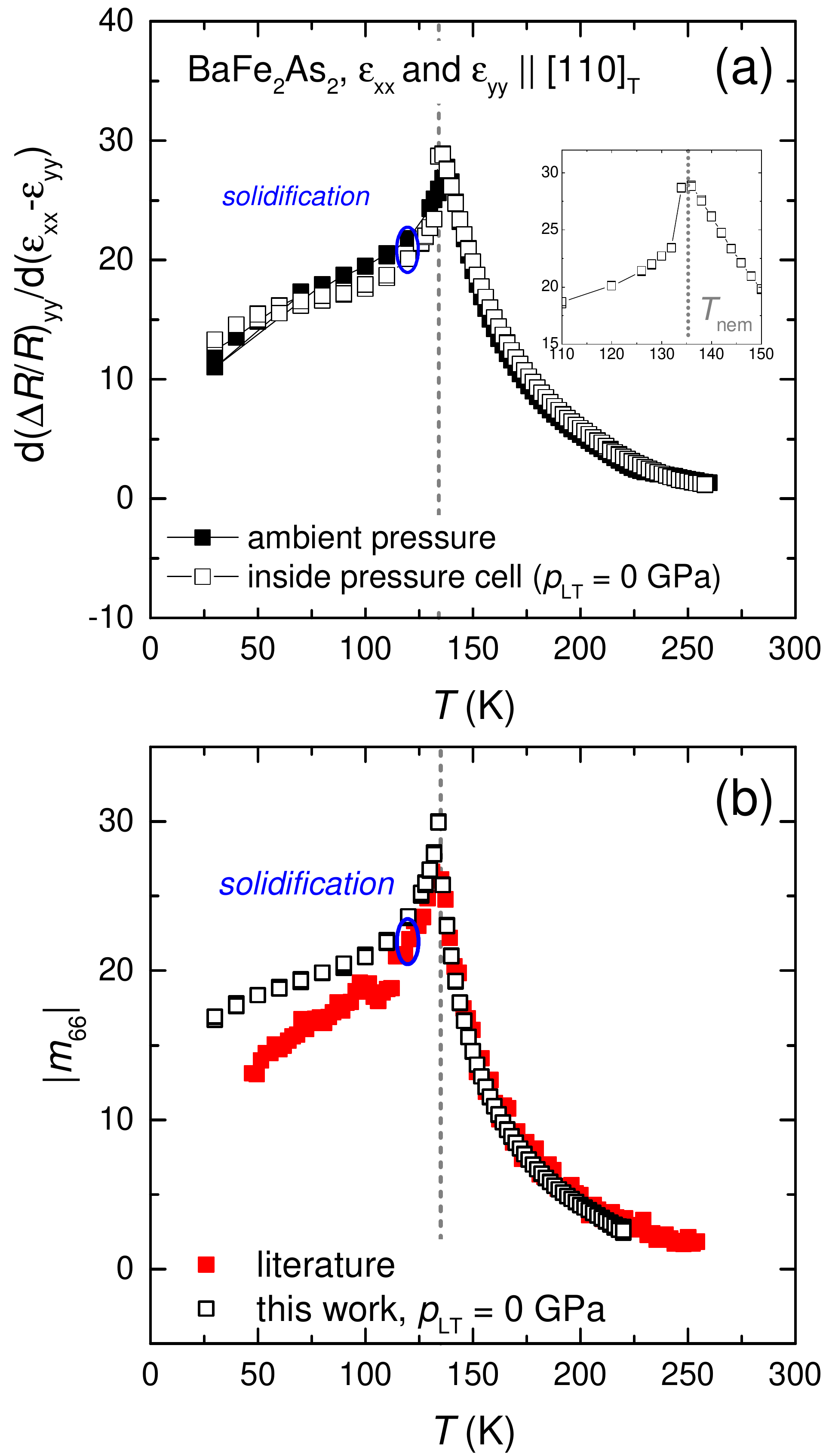}
	\caption{(a) Comparison of transverse elastoresistance, d$(\Delta R/R)_{yy}$/d$(\epsilon_{xx}-\epsilon_{yy})$, of BaFe$_2$As$_2$, taken at ambient pressure (outside of the pressure cell) and inside the pressure cell at lowest pressure, corresponding to a pressure of $p_{LT}\,=\,$0\,GPa for low temperatures (determined by the Pb manometer); (b) Comparison of the elastoresistance coefficient, $|m_{66}|$, (obtained from ($(\Delta R/R)_{xx}-(\Delta R/R)_{yy}$)/d$(\epsilon_{xx}-\epsilon_{yy})$), see main text for details) of BaFe$_2$As$_2$ between published data in literature at ambient pressure \cite{Kuo16} and data, obtained in the present work inside the pressure cell at lowest pressure ($p_{LT}\,=\,$0\,GPa). Dashed line indicates the position of the nematic transition at $T_{nem}$ in both panels, as shown on enlarged scales in the inset of (a). The blue circle marks the position of the solidification of the pressure medium for this particular pressure. Our data in (b) is only plotted up to 220\,K due to non-linearities in the $R_{xx}$ vs. $\epsilon$ data above 220\,K.}
	\label{fig:Ba122-invsout} 
	\end{figure}
	\end{center}

	From the slope of a linear fit to the experimental data in Fig.\,\ref{fig:Ba122-Rvseps}, the elastoresistance coefficients are obtained. First, we compare measurements of elastoresistance, taken at ambient pressure (i.e., outside of the pressure cell, but mounted on the pressure-cell feedthrough), with those, taken inside the pressure cell at lowest pressure (corresponding to a pressure of 0\,GPa at low temperature, as determined from the Pb manometer). The elastoresistance coefficients d$N$/d$\epsilon$\,=\,d$(\Delta R/R)_{yy}$/d$(\epsilon_{xx}-\epsilon_{yy})$, obtained in these two conditions, are shown as a function of temperature, $T$, in Fig.\,\ref{fig:Ba122-invsout}\,(a). At $T\,=\,260\,$K, d$N$/d$\epsilon\,\approx\,1$, which is a typical value for materials, in which geometric effects rather than intrinsic resistance changes (e.g. due to nematicity) dominate (for comparison, the gauge factor, d$(\Delta R/R)$/d$\epsilon$, for constantan is $\approx\,$2). The sign of d$N$/d$\epsilon$ over the full temperature range is consistent with previous reports for the transverse resistance from elastoresistance measurements \cite{Kuo16}, as well as with the resistivity anisotropy, measured in the orthorhombic state using a mechanical clamp \cite{Chu10,Blomberg12}. Upon cooling from 260\,K,  d$N$/d$\epsilon$ increases strongly and peaks at $T\,\approx\,$135\,K at a value of $\approx\,30$. This temperature corresponds to the nematic transition temperature $T_{nem}$ (i.e., structural transition temperature) of BaFe$_2$As$_2$ at ambient pressure, at which the tetragonal-to-orthorhombic phase transition occurs \cite{Kim11,Rotter08,Ni08}. Below $T_{nem}$,  $|$d$N$/d$\epsilon |$ drops suddenly, and exhibits a much weaker $T$-dependence. The large values of d$N$/d$\epsilon\,\gg\,2$, together with the observation of a peak just at $T_{nem }(p\,=\,0)$ strongly suggest that the elastoresistance of BaFe$_2$As$_2$ is dominated by contributions of nematic fluctuations for $T\,>\,T_{nem}$. In contrast, in the low-temperature, nematically-ordered state, the elastoresistance is dominated by domain formation, associated with the tetragonal-to-orthorhombic distortion and disorder, and therefore is non-universal. Indeed, the two data sets of d$N$/d$\epsilon$, taken inside and outside the pressure cell, show an excellent agreement for $T\,>\,T_{nem}$ (defined by the maximum position of d$N$/d$\epsilon$, see below), but slight discrepancies for $T\,<\,T_{nem}$. This can therefore be taken as a strong indication that the pressure cell environment, which gives rise to small pressure changes with temperature, does not alter the measurements of d$N$/d$\epsilon$.
	
	Furthermore, these data can be compared to those, reported in literature \cite{Kuo16} for BaFe$_2$As$_2$. As the literature data were obtained using a different actuator with different $\nu_p\,\approx\,2$ ($\neq 1$), a full symmetry decomposition has to be performed here by subtracting the response of two samples, i.e., d$N$/d$\epsilon$\,=$\,-2\,m_{66}\,=\,$d$[(\Delta R/R)_{xx}-(\Delta R/R)_{yy}]$/d$(\epsilon_{xx}-\epsilon_{yy})$. The so-derived $m_{66}$ values from the present study (inside the pressure cell) and the literature values are presented in Fig.\,\ref{fig:Ba122-invsout}\,(b). Again, whereas the behavior for $T\,<\,T_{nem}$ is clearly (but not significantly) different, the data for $T\,>\,T_{nem}$ show a very good agreement in the $T$ dependence and the absolute values.

\begin{center}
	\begin{figure}
	\includegraphics[width=0.5\textwidth]{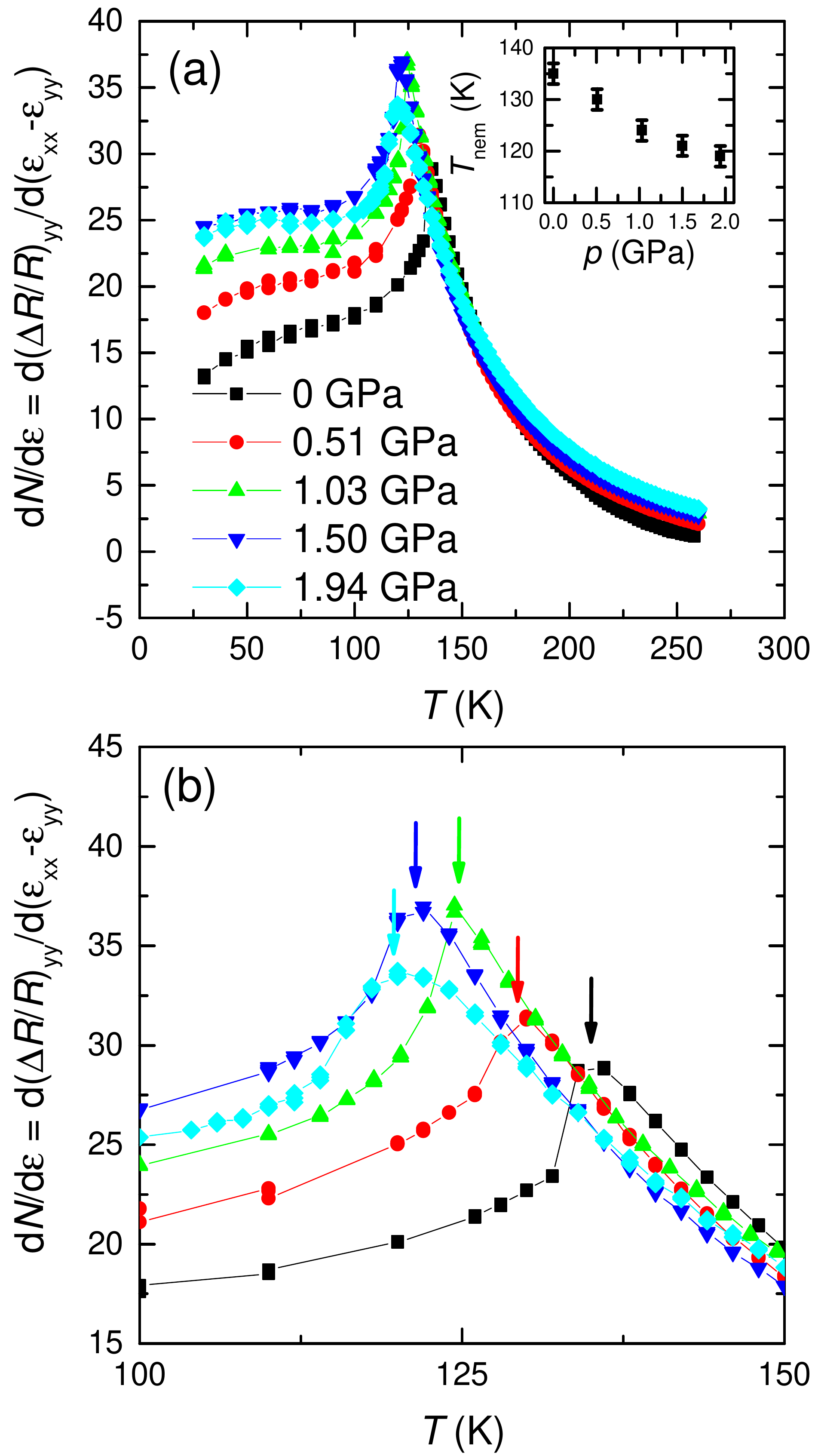}
	\caption{Transverse elastoresistance, d$(\Delta R/R)_{yy}$/d$(\epsilon_{xx}-\epsilon_{yy})\,=\,$d$N$/d$\epsilon$, of BaFe$_2$As$_2$ as a function of temperature, $T$, for different pressures between 0\,GPa and 1.94\,GPa (a). Panel (b) shows the data, shown in (a) on enlarged scales around the nematic transition at $T_{nem}$, indicated by the arrows for each pressure. The inset in (a) depicts the change of $T_{nem}$ with pressure, $p$.}
	\label{fig:Ba122-pressure} 
	\end{figure}
	\end{center}

	Since our d$N$/d$\epsilon$ data inside the pressure cell are in good agreement with our own measurements outside of the cell, as well as with the literature, we can now proceed to discuss the effect of pressure on d$N$/d$\epsilon$. The results for pressures up to 1.94\,GPa are shown in Fig.\,\ref{fig:Ba122-pressure}\,(a) over the full $T$ range, as well as in 	Fig.\,\ref{fig:Ba122-pressure}\,(b) on expanded scales around $T_{nem} (p)$. Overall, the form of d$N$/d$\epsilon$\,=\,d$(\Delta R/R)_{yy}$/d$(\epsilon_{xx}-\epsilon_{yy})$ is not significantly affected by pressure; d$N$/d$\epsilon$ is strongly $T$-dependent for $T\,>\,T_{nem}(p)$, shows a drop when cooling through $T_{nem}(p)$ and is only weakly $T$-dependent for $T\,<\,T_{nem}(p)$. A sizable change, however, can be resolved in the position of the peak in temperature, which shifts to lower temperatures with increasing $p$. To quantify the shift with $p$, we estimate the position of the maximum in d$N$/d$\epsilon$ for each pressure and use it as a proxy for $T_{nem}$, the precise determination of which is somewhat limited by the $T$ spacing ($\Delta T\,=\,2$\,K) of the data sets. The so-derived $T_{nem}$ decreases monotonically with pressure, at a rate of d$T_{nem}$/d$p\,\approx\,-(8.5\,\pm\,1)$\,K/GPa. This suppression rate agrees very well on a quantitative level with our recent thermodynamic studies \cite{Gati19} of $T_{nem}(p)$ (d$T_{nem}$/d$p\,\approx\,-9$\,K/GPa). This agreement might be considered as an indication that the sample is pressurized in an almost hydrostatic manner (or in other words, the in-plane compressibility mismatch between sample and actuator appears to be small). However, given that purely $c$ axis uniaxial pressure is also expected \cite{Meingast12} to shift the nematic transition temperature to lower temperatures with increasing $p$, a dominant $c$ axis uniaxial contribution cannot be ruled out at present, even though the quantitatively similar values of d$T_{nem}$/d$p$ rather point towards a more hydrostatic pressure environment.

\begin{center}
	\begin{figure}
	\includegraphics[width=\textwidth]{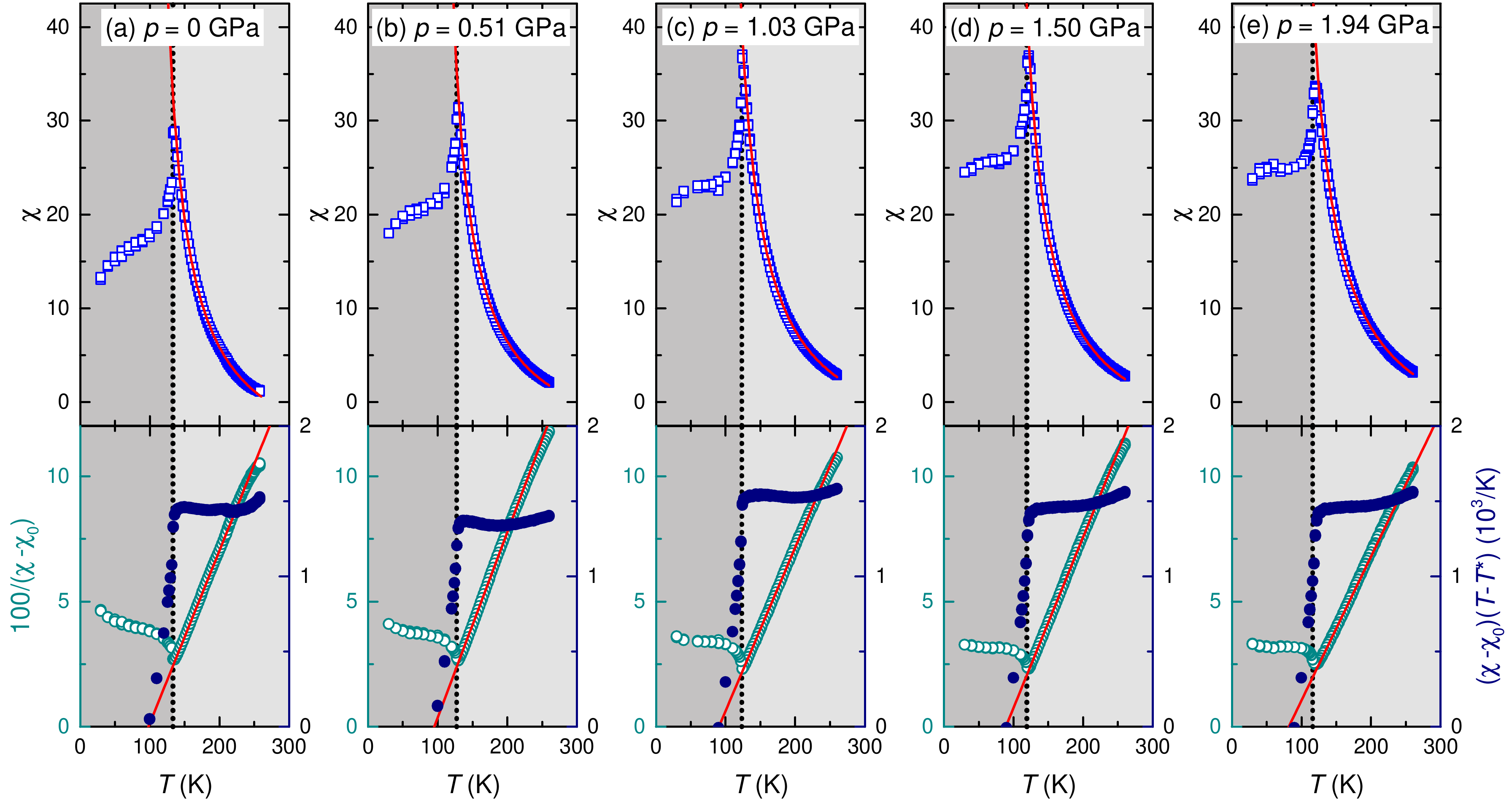}
	\caption{Temperature ($T$) dependence of nematic susceptibility, $\chi_{nem}$, of BaFe$_2$As$_2$ for pressures between 0\,GPa and 1.94\,GPa (a)-(e) (for definition of $\chi_{nem}$, see main text). In each panel, the top plot shows the experimentally-determined temperature dependence of $\chi$ (open blue symbols), together with a Curie-Weiss fit, $\chi\,=\,\chi_{nem}+\chi_{0}$, (red line) to the data above the nematic transition temperature $T_{nem}$ (light grey area). The bottom plot shows the same data, represented as the inverse nematic susceptibility, 1/$(\chi-\chi_{0})$, (left axis; open symbols represent the experimental data, red line the Curie-Weiss fit) and as the Curie constant $C$, corresponding to $(\chi-\chi_{0})(T-T^{\star})$. $\chi_{0}$ describes the intrinsic elastoresistance effect, which does not originate from nematicity and describes geometric effects, and $T^{\star}$ corresponds to the Curie temperature. The procedure, which is used to determine these two parameters, is described in the main text.}
	\label{fig:Ba122-Curiefitting} 
	\end{figure}
	\end{center}

\begin{center}
	\begin{figure}
	\includegraphics[width=0.8\textwidth]{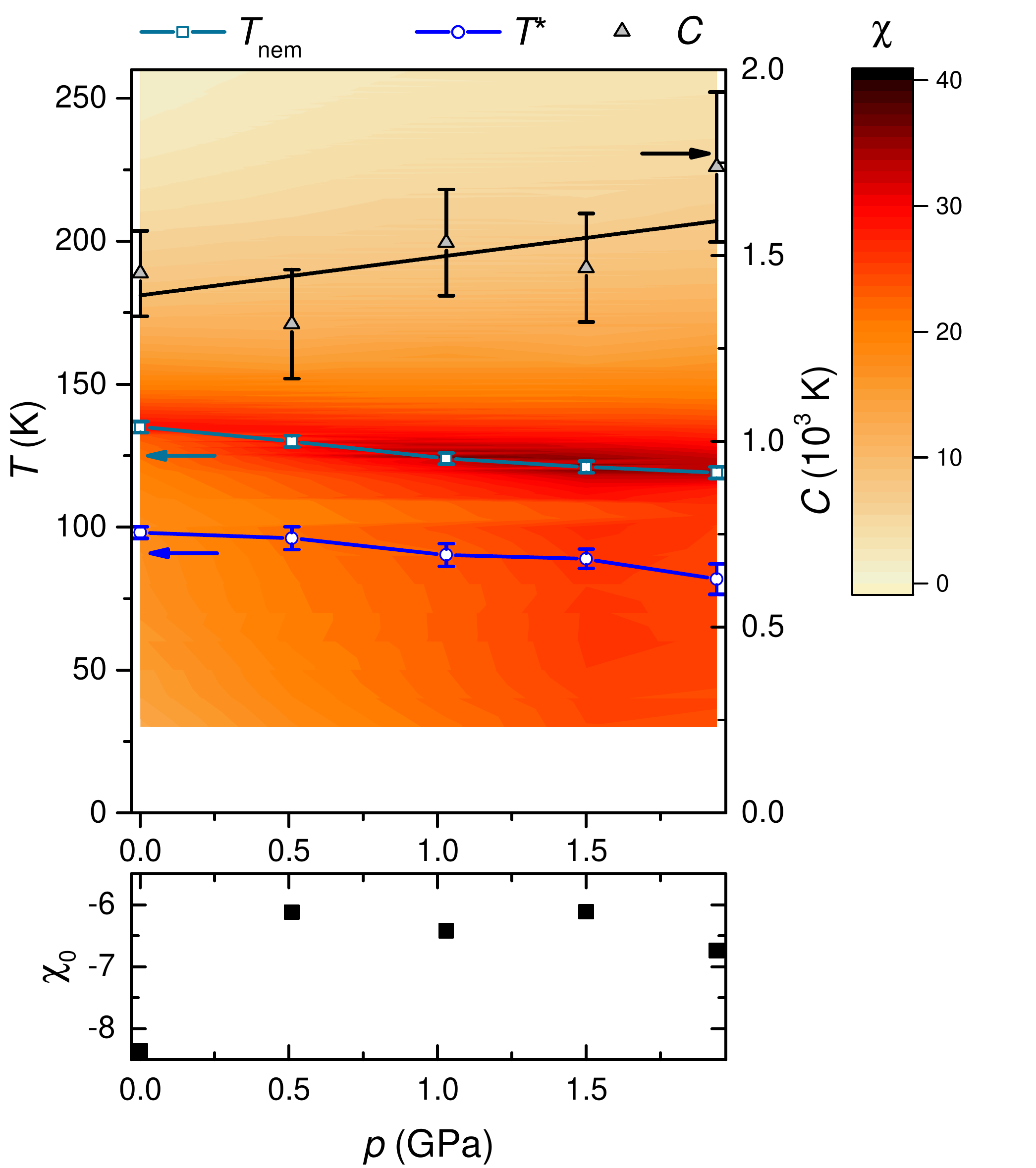}
	\caption{Evolution of nematic transition temperature, $T_{nem}$ (cyan, open squares), Curie-Weiss temperature, $T^\star$ (open, blue circles), (left axis) and Curie constant, $C$ (grey triangles), (right axis) with pressure, $p$. The latter two quantities were extracted from the fits, shown in Fig. \ref{fig:Ba122-Curiefitting}. The color profile in the background represents the magnitude of $\chi$. The black line, which connects the $C$ data points, is a guide to the eye. The bottom panel depicts the evolution of the temperature-independent fitting parameter $\chi_0$ with pressure.}
	\label{fig:Ba122-CWparameters} 
	\end{figure}
	\end{center}

	The size and temperature dependence of d$N$/d$\epsilon$ for $T\,>\,T_{nem}$ reveal information on the nematic fluctuations. For example, for $T\,=\,140$\,K, the size of d$N$/d$\epsilon$ decreases monotonically with $p$, whereas for 260\,K, d$N$/d$\epsilon$ has opposite behavior. In general, the nematic susceptibility, $\chi_{nem}$, is expected to diverge when approaching the bare (i.e., without coupling to the crystal lattice) electronic nematic transition temperature $T^\star$. The coupling of the electronic subsystem to the lattice induces the structural phase transition at $T_{nem}$ and also, raises the transition temperature such that $T_{nem}\,>\,T^\star$. The minimization of a free energy expansion (including a symmetry-allowed bilinear coupling term between the lattice strain $\epsilon$ and the electronic nematic order parameter $\psi$) \cite{Kuo13}, yields that $\chi_{nem}$ should follow a Curie-Weiss-like behavior,
	
	\begin{equation}
	\chi_{nem}\,=\,\frac{C}{T-T^\star},
	\end{equation}

with $C$ being directly proportional to the strength of the bilinear coupling.  In the following, we discuss a fitting of our data with this Curie-Weiss ansatz (following the original works of Ref.\,\citen{Chu12,Kuo13,Kuo16}). Experimentally, $\chi_{nem}$ can be approximated by the elastoresistance coefficient, which is related to nematicity (see Sec. \ref{sec:results-Ba122-intro} above). In addition, a temperature-independent contribution $\chi_0$, which is unrelated to electronic nematicity, has to be considered. As a result, d$N$/d$\epsilon\,\simeq\,\chi\,=\,\chi_{nem}+\chi_{0}$, with d$N$/d$\epsilon\,=\,$d($(\Delta R)/R)_{yy}$/d$(\epsilon_{xx}-\epsilon_{yy})$. During the fitting procedure, care has to be taken in choosing the temperature range for the fit; typically, deviations can be expected for very high temperatures ($T\,>\,250\,$K) due to a progressive decrease of strain transmission through the epoxy with higher temperatures. In addition, in earlier works, deviations close to $T_{nem}$ were reported and attributed to effects of disorder. To determine the fitting range, we followed the procedure, described in Ref.\,\citen{Kuo16}. The typical fitting range in the present case was chosen to be $(T_{nem}+4\,$K$)\,\le\,T\,\le\,250\,$K. Error bars of the fitting parameters are estimated from their variation upon performing fits in in total four different temperature windows. The result of the fit is shown in the top panel of Fig.\,\ref{fig:Ba122-Curiefitting} for all five pressures (with $-8\,\lesssim\,\chi_0\,\lesssim\,-6$ for all data sets, see bottom panel in Fig.\,\ref{fig:Ba122-CWparameters}). For each pressure, the data for $T\,>\,T_{nem}$ (light grey regime in Fig.\,\ref{fig:Ba122-Curiefitting}) can be well described with a Curie-Weiss-like behavior. The quality of the fit is better visible in plots of the inverse nematic susceptibility $1/\chi_{nem}\,=\,1/(\chi-\chi_0)$ and a plot of the Curie constant, given by $(\chi-\chi_0)(T-T^\star)$ (left and right axis of the bottom panel of Fig.\,\ref{fig:Ba122-Curiefitting}, respectively). These representations clearly demonstrate the quality of the fit, as only small deviations can be observed for higher temperatures ($T\,\gtrsim\,240\,$K), likely due to above-outlined issues with the epoxy.

	The observation of a $1/(T-T^\star)$ dependence of the elastoresistance (i.e., of the nematic susceptibilty) for all pressures up to $\approx\,2\,$GPa speaks in strong favor of nematic electronic fluctuations \cite{Fernandes14}, which drive the structural distortion in BaFe$_2$As$_2$ in this pressure range. Further information on nematicity under pressure can be inferred from the pressure dependence of the fitting parameters $T^\star$ and $C$, which are presented in Fig.\,\ref{fig:Ba122-CWparameters}, together with $T_{nem}(p)$. The temperature of the bare electronic transition $T^\star$ is lower than $T_{nem}(p)$ for all pressures. For ambient pressure, this result was observed in previous studies \cite{Kuo16,Boehmer14} and can be rationalized with the free energy ansatz, presented above. In fact, we find that $T^\star$ is suppressed at a very similar rate as $T_{nem}$ is, which implies that $T_{nem}-T^\star\,\approx\,$const. up to $\approx\,$2\,GPa. At the same time, the Curie constant $C$, which characterizes the strength of electron-lattice coupling is, within the error bars, almost constant, with a minimal tendency towards increasing $C$ with $p$ (d$C$/d$p\,\approx\,(100\,\pm\,30)$\,K/GPa). All together, whereas our results clearly show that nematic fluctuations prevail, they also suggest that the coupling strength between nematic order and the lattice (measured by $T_{nem}-T^\star$ and $C$, respectively) is essentially unaffected by pressures up to 2\,GPa. Similar observations were made in investigations on samples of BaFe$_2$As$_2$ with small Co doping or K doping from a variety of probes \cite{Kuo16,Boehmer14,Boehmer16}, which reveal information on the nematic susceptibility. For higher doping levels, close to optimal doping, measurements of the elastoresistance revealed ubiquitous features of nematic quantum criticality in various members of the iron-based family \cite{Kuo16}. In light of these results, it remains to be seen in the future, whether the application of pressure in the BaFe$_2$As$_2$ family can also reveal signatures of nematic quantum criticality, similar to those associated with optimal doping. The results of elastoresistance under pressure on pure BaFe$_2$As$_2$, presented here, show that such questions can be addressed with the present setup by tuning a single sample through the possible quantum critical point.

\section{Summary}

The present work describes the experimental realization of a setup to tune materials by uniaxial stress and hydrostatic pressure simultaneously. The design utilizes a piezoelectric actuator, which allows for the change of quasi-uniaxial stress \textit{in situ} inside the piston-cylinder pressure cell. By carefully characterizing the expansion of the piezoelectric actuator upon application of an external voltage, we show that the actuator reliably operates in the full temperature and pressure range (below room temperature and below 2\,GPa, respectively). In addition, we present measurements of the resistance of the iron-based compound BaFe$_2$As$_2$ as a function of strain (i.e., measurements of the elastoresistance) for finite pressures as a proof-of-principle example. We demonstrate (i) that the amount of strain, generated in our setup by the stress of the actuator, is sufficient to unravel a sample response, and (ii) that the sample response can be modeled in terms of nematicity, omnipresent in this material family. Therefore, the present setup is able to address interesting scientific questions in the study of novel electronic states. It also points out that this particular combination of symmetry- and non-symmetry-breaking tuning parameters is experimentally realizable and will therefore motivate further optimization and advancements.

	We again want to stress that the present work serves to demonstrate that piezoelectric actuators can be operated inside piston-cylinder pressure cells to apply uniaxial stress \textit{in situ}. Future technical works shall certainly involve improvements of hydrostaticity of the pressure environment, but might therefore also involve the implementation of other measurement techniques, tuning parameters (such as magnetic field\cite{Jo19}) or the generation of larger strain to make best use of this capability. 


\begin{acknowledgments}
We thank N. Ni and A. Thaler for growing the BaFe$_2$As$_2$ crystals used in the study, and R. Valent\'{i} for discussions that lead to this research direction. Work at the Ames Laboratory was supported by the US Department of Energy, Office of Science, Basic Energy Sciences, Materials Sciences and Engineering Division. The Ames Laboratory is operated for the US Department of Energy by Iowa State University under Contract No. DEAC02-07CH11358. E.G. and L.X. were funded, in part, by the Gordon and Betty Moore Foundation's EPiQS Initiative through Grant No. GBMF4411. 
\end{acknowledgments}

\bibliographystyle{modaps}

\end{document}